\documentclass[prl,twocolumn,showpacs,floatfix,amssymb]{revtex4}
\usepackage{graphicx}

\begin{document}
\title{Observation of Convergent Oscillations of the Flux Line Lattice
as a Result of Magnetic Flux Jumping in Hard Superconductors}
\author{Victor V. \surname{Chabanenko}}
\affiliation{Physico-Technical Institute,  National Academy of
Sciences, 72 R. Luxemburg str., 83114 Donetsk, Ukraine}
\author{Vladimir F. \surname{Rusakov}}
\affiliation{Donetsk National University,
24 Universitetskaja str., 83055 Donetsk, Ukraine}
\author{Valery A. \surname{Yampol'skii}}
\affiliation{Institute for Radiophysics and Electronics, NAS, 12 Acad. Proskura str., 61085 Kharkov,
Ukraine}
\author{Stanislaw \surname{Piechota}} \author{Adam \surname{Nabialek}}
\author{Sergei V. \surname{Vasiliev}} \author{Henryk \surname{Szymczak}}
\affiliation{Instytut Fizyki PAN, 32 Al. Lotnik\'{o}w, 02-668
Warszawa, Poland}

\date{\today}

\begin{abstract}
We have monitored new peculiarities of the dynamics of
catastrophic avalanches of the magnetic flux in superconducting
Nb, Nb-Ti, and YBaCuO samples: i) convergent oscillations of the
magnetic flux; ii) a threshold for entering the huge flux
avalanches in the shielding experiments; iii) a threshold for the
exit of a residual flux in the trapping experiments. The observed phenomena are interpreted in
terms of a theoretical model which takes into account the inertial
properties of the vortex matter.
\end{abstract}
\maketitle

Recent studies of the marginally stable Bean critical state determined by a balance
between intervortex repulsion and pinning by defects in type II superconductors have
brought out a number of astonishing phenomena associated with spatiotemporal dynamic
properties, vortex avalanches of many scales, self-organized criticality and memory
effects in vortex matter (for example, \cite{Field}).

Our experiments identify flux avalanches as nonlocal processes
propagating rapidly over the distance comparable with the size of
the sample. Earlier thermal effects in massive rods of niobium
caused by giant flux jumps were observed in \cite{Zebouni}. Some
observations of the behavior of the flux jumps and its structure
were presented in \cite{Leblank}. These data show that there
exists the exact correlation between the jumps in the resistivity
and the collapse of magnetizaton. One of the
recent investigations \cite{Geim} of magnetic field screening by
the superconductors in the vortex state revealed "negative
vortices" whose penetration leads to the expulsion of a magnetic
field.

In this paper we report about the observation of two distinct
magnetic phenomena in the vortex state of superconductors: i) a threshold
for entering the huge flux avalanches in the shielding experiments
(in other words, the increase in the screening properties of a
superconductor before the flux jump). We have also found a
threshold for the exit of a residual flux in the trapping
experiments; ii) convergent oscillations of the magnetic flux
arising from flux jumping.

The latter result is of a particular interest since collective modes in a fluxoid system
are very difficult to observe \cite{Suhl}. In addition, due to a large vortex viscosity,
the displacement waves in a vortex lattice do not propagate.

The observed oscillations in the vortex system can be interpreted
as a result of the existence of a definite value of the effective
vortex mass, i.e. this phenomenon can be considered as displaying
the inertial property of the vortex matter.

We have studied the structure of the flux jumps by means of a miniature Hall probe sensor.
It was placed in the center of the sample (Fig.~1) and measured the surface induction
$B_0$. We have examined the Hall sensor voltage directly by the transient recorder (model
TCC-1000, Riken Denshi Co., LTD). In our experiment a flux avalanches are detected in real
time ($t_{real}>10^{-6}$~s) without distortion. In the shielding experiments, the zero
field cool mode (ZFC) was used. The trap field mode was implemented by increasing the
external magnetic field up to the values exceeding the critical magnetic field $H_{c2}$.
The data presented in this study were obtained from Nb-Ti(at\% 50) samples,
polycrystalline Nb plates, and melt-textured YBaCuO slabs. The sizes ($\mathrm{L\times
l\times 2R\:mm^3}$, see Fig.~1) of the samples were as follows: for Nb-Ti, $15\times
2.8\times 3.1\:\mathrm{mm}^3$; for Nb, $11\times 3.5\times 5\:\mathrm{mm}^3$; for YBaCuO,
$6\times 2\times 3.5\:\mathrm{mm}^3$.

\begin{figure}
\includegraphics{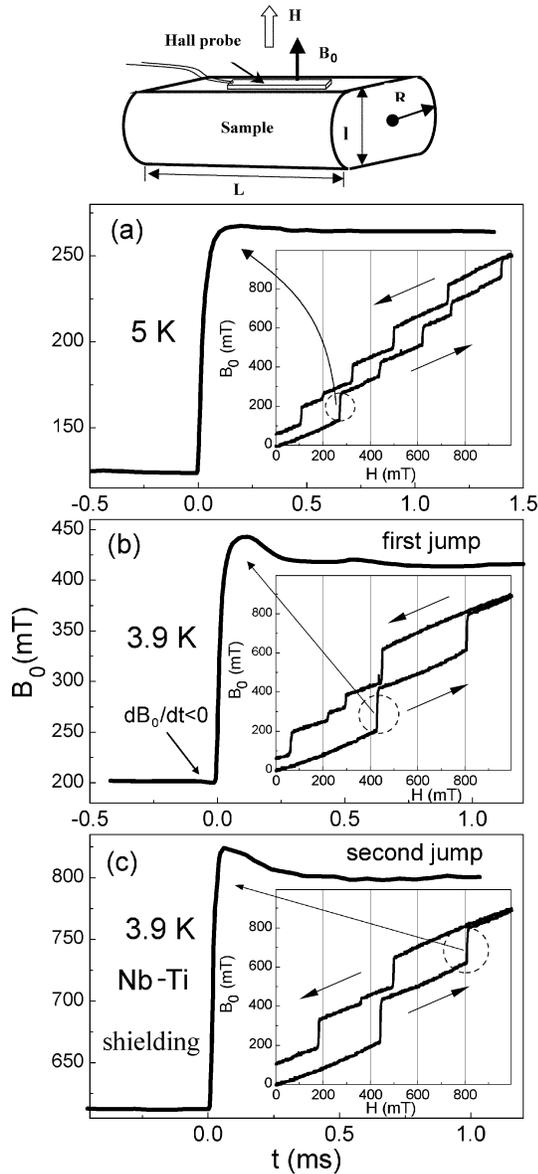}
\caption{Above - geometry of the experiment. Main frames: the temporal evolution of the
surface magnetic induction $B_0$ for the first flux jump (a, b) and for the second flux
jump (c). Inserts: Hysteresis loops, $B(H)$} \label{fig1}
\end{figure}

Fig.~1(a-c) shows the temperature and temporal dependence of the surface magnetic induction
$B_0$ for the first and the second flux jumps in a shielding mode for the Nb-Ti sample.
The structure of the jump essentially differs from a simple step. Some of the principal
differences are clearly seen in Fig.~2(a).

The development of the thermomagnetic instability can be divided into three stages (see
Fig.~1(b) and Fig.~2(a)). The flux jump is preceded by a certain phenomenon under which
the field on a superconductor surface decreases (see Fig.~2(a) for Nb-Ti, Fig.~3(a) for
Nb, and Fig.~4 for YBaCuO). The value of the negative peak in the experiment with Nb is
16\% of the total value of the flux jump. This is the first stage of the process. The
second stage is in fact the avalanche-like penetration of the flux. Finally, at the last
stage the relaxation of thermal and conductive properties of the superconductor occurs. In
the most cases, this is a nonmonotonic process. Let us analyze it for the Nb-Ti sample.
\begin{figure}
\includegraphics{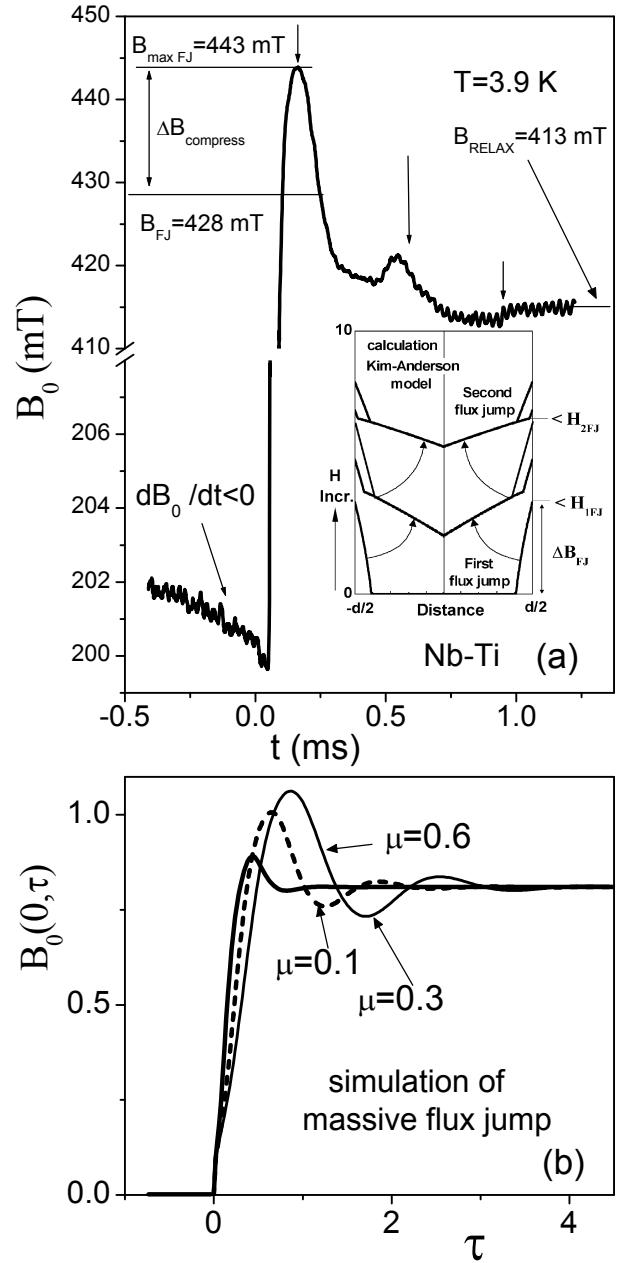}
\caption{(a) The fine structure of the first jump for the Nb-Ti sample; (b) The results of
simulation of the oscillation process in the vortex system; insert: distribution $B$ vs
distance $x$ for the first and the second jumps} \label{fig2}
\end{figure}

As is shown in Fig.~2(a), after the avalanche had penetrated the sample, the induction
$B_0(t)$ in the center of the sample was higher than that of the external magnetic field
by $\Delta B_{comp}$. Recently it was found by Nowak \cite{Nowak}, that the large
avalanches are most likely system spanning; as a result after avalanche a significant
fraction of the Nb film has magnetic induction large than the applied field. Apparently,
Coffey \cite{Coffey} observed such a compression of the vortex lattice by measuring the
distribution of the induction inside the sample after the flux jump. Next, in the
background of a relaxation curve the damped oscillations with a characteristic frequency
$\nu \simeq 2.5$~kHz were observed.

The effect of the compression of the magnetic flux caused by avalanche passing is
particularly pronounced in the YBaCuO samples as well. In order to check this effect we
measured the difference $B_0-\mu_0 H$ using two Hall probes where $H$ is the external
magnetic field, $\mu_{0}$ is the permeability of vacuum. Fig.~4 shows that this difference
changes its sign after the avalanche has passed. The difference $\Delta B_{comp}$ is about
the value of the penetration field $H_p$.

Now, let us consider peculiarities of the structure of the thermomagnetic avalanche for
the case of magnetic flux trapping. For Nb, (Fig.~3(b)) a positive peak $B_0(t)$ is
observed which precedes the development of the instability. It is of the same origin as
the negative peak $B_0(t)$ in the case of the shielding mode (Fig.~3(a)).
\begin{figure}
\includegraphics{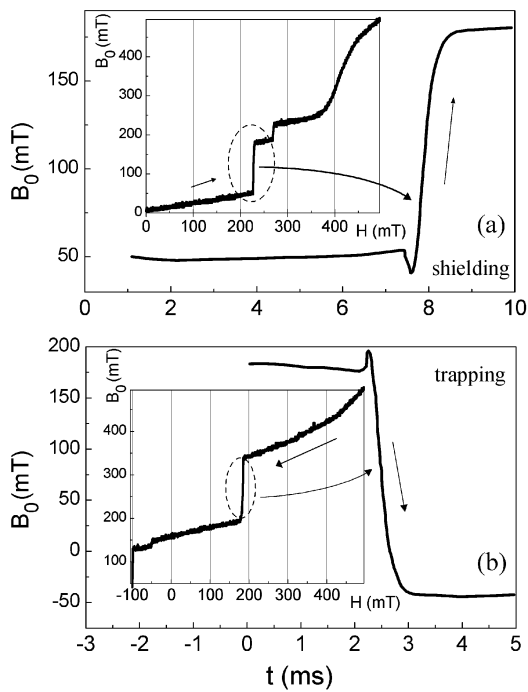}
\caption{Structure of the avalanche jumps for (a) shielding and (b) trapping regimes for
Nb. Inserts show $B_0(H)$.} \label{fig3}
\end{figure}

The above-described peculiarities of the flux jumps are not
sensitive to the change in the rate ${\rm d}H/{\rm d}t$ of the
external magnetic field sweeping in the range of $0.05-2$~T/s.

Now we would like to scrutinize the observed phenomena. In our opinion the most striking
effect is the compression of the magnetic flux after the avalanche has passed through the
sample. It is very important to emphasize that the magnetic induction in the middle of the
sample can exceed the value of the external magnetic field. In this case, some of the
vortices move in the direction from the area with low vortex density towards the area with
the vortex surplus, against the total repulsion force. This means that these vortices move
in spite of all the forces acting in the direction opposite to this motion. Such a motion
is possible only due to the existence of some effective vortex mass. There are plenty of
papers dealing with the theoretical attempts to estimate the mass of the Abrikosov vortex
\cite{Suhl,Kopnin}. Experimental data on microwave field attenuation contain information
on the inertial properties of a vortex \cite{Rose}.
\begin{figure}
\includegraphics{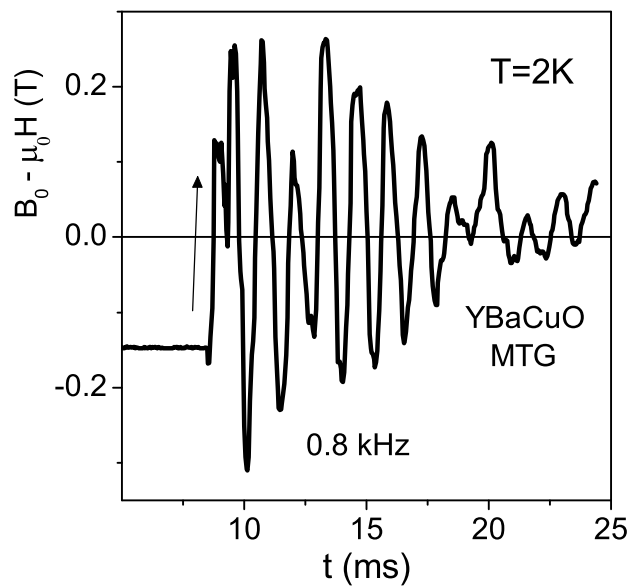}
\caption{The temporal dependence of the difference of magnetic induction in the middle of
the sample $B_0$ and the external field $\mu_0 H$ for the YBaCuO plate.}\label{fig4}
\end{figure}

To explain the observed oscillatory phenomena, we suggest the following mechanism which
takes into account the inertial properties of the vortex system. Prior to the jump, the
mixed state of superconductors is characterized by nonuniformly distributed magnetic
induction localized near the surface. As a result of the avalanche, the flux
rushes from either sides of the sample towards the center. Two fronts of the penetrating
flux collide in the center of the sample and, owing to the existing vortex mass, give
rise to the local surplus density of the magnetic flux that exceeds the value of the
external magnetic field. The repulsion force in the vortex structure at the center of the
sample that have resulted from its compaction, initiates the wave of the vortex density
of the inverse direction of propagation. Upon reaching the surface, this wave is
reflected from it. This results in the oscillations in the vortex system. The limitation
of the number of oscillations observed is caused by the existence of damping. One
succeeds in observing the oscillation of the vortex density only owing to a strong
compression of the vortex structure as a result of the giant avalanche-flux. But for the
second jump (Fig.~1(c)), a smooth descending slope in the curve $B_0~(t)$ is noted and
the oscillatory mode is not observed.

The observed process of damping oscillations can be described in
terms of the following theoretical model. We use the continuity
equation for the dimensionless variables,
\begin{equation}
\dot{b}+(bv)^{\prime}=0 \label{1}
\end{equation}
where $b=B/H_p=B c/2\pi J_c(0)d$ is the normalized magnetic
induction, $d$ is the sample thickness, $J_c(0)$ is the critical
current density at the initial temperature (before the avalanche
has passed), the dot denotes the derivative with respect to the
time $\tau=t/t_0=t\Phi_0 H_p/4\pi d^2\eta$, $\Phi_0$ is the flux
quantum, $\eta$ is the viscosity coefficient, the sign prime
denotes the derivative with respect to the coordinate $\xi=x/d$,
$v=V/V_0=Vt_0/d$ is the hydrodynamic vortex velocity. The second
equation of the model describes the motion of the vortex system
under the action of the Lorentz, pinning, and viscosity forces,

\begin{equation}
\mu v'v+\mu \dot{v}=-b'-2 \rm{sign}(v) f(\theta)-v. \label{2}
\end{equation}
Here $\mu=m/m_0=m\Phi_0 H_p/4\pi d^2\eta^2$ is the normalized
vortex mass, $\theta=4\pi C \Delta T/H^2_p$, $C$ is the specific
heat, $\Delta T=T-T_0$ is the deviation of the temperature from
its initial value $T_0$ before the flux jump, the function
$f(\theta)$ describes the temperature dependence of the critical
current, $J_c(\theta)=J_c(0) f(\theta)$, $f(\theta)=1$ at $\theta
=0$ and $f(\theta)=0$ at $\theta =T_c-T_0$ ($T_c$ is the critical
temperature). The temperature change occurs due to the Joule
losses. Neglecting the process of thermal conductivity, we can use
the following equation of the energy balance:
\begin{equation}
\dot{\theta}=b v^{2}+2|v| b f(\theta). \label{3}
\end{equation}

This set of equations should be solved within the spatial interval
$-1/2 <\xi <1/2$ together with the boundary and initial
conditions, $b(1/2,\tau)=b(-1/2,\tau)=H/H_p=\alpha, \quad
b(\xi,0)=\alpha(1+2(\beta/\alpha)(|\xi|-1/2)), \quad
\theta(\xi,0)=0, \quad v(\xi,0)=0.$ Here $\beta=J_0/J_c$, $J_0$ is
the critical current density at the initial time moment. We assume
that this value exceeds slightly the value $J_c(0)$, i.e. $\beta
>1$. This supposition is introduced to initiate the onset of an
avalanche.

The results of the simulation for $\alpha=1$, $\beta=1.01$, $f(\theta)=1-\kappa \theta$,
$\kappa=H_p^2 /4\pi C(T_c -T_0)=12$ are presented in Fig.~2(b). It is seen that the
curves in this figure describe qualitatively well the behavior of the experimental ones
(see Fig.~2(a)).

The suggested mechanism of the oscillations which takes into account the inertial
properties of the vortices allows one to interpret the surprising behavior of the vortex
system just before the avalanche passage when the magnetic flux near the middle of the
sample unexpectedly moves in the opposite-to-avalanche direction (see Figs.~3(a),(b)).
The mechanism of this effect may be as follows. Due to the vortex mass, the frozen
profile of the magnetic induction after some avalanche has passed can contain a smooth
peak in the middle of the sample (see, for example, \cite{Coffey}). This local maximum
remains on hand while the external field increases and exists in the initial profile for
the subsequent avalanche. The new avalanche can destroy this peak due to the heating
effect before the new vortices arrive there. This tends to decrease the vortex density in
the middle of the sample.

Finally, we remark that very interesting oscillating phenomena
originated by the thermomagnetic avalanches are observed in the
vortex matter for both low-$T_c$ and high-$T_c$ hard
superconductors. These phenomena can be qualitatively interpreted
in terms of the theoretical model which takes into account the
inertial properties of the vortex system. However, new
experimental and theoretical in-depth studies are necessary to
elucidate the situation.

\begin{acknowledgments}
We are greateful to V. Vinokur, J. Kolachek, A.E. Filippov, V.A. Shklovskij, Yu. Genenko,
A.I. D'yachenko, and A. Abal'oshev for the discussions of this problem. The work was
partly supported by the Polish Committee for Scientific Research (grants No 8 T11B 038
17).
\end{acknowledgments}


\begin{thebibliography}{100}
\bibitem{Field} S. Field, J. Witt, F. Nori, X. Ling, Phys. Rev. Lett. {\bf 74},
1206 (1995); R.J. Zieve et al., Phys.Rev.B. {\bf 53}, 11849 (1996); K. Behnia, C. Capan, D. Mailly, B. Etienne,
Phys.Rev.B. {\bf 61}, R3815 (2000); G.T. Seidler, et al., Phys.Rev.Lett. {\bf
74}, 1442 (1995); K.E. Bassler, M. Paczuski, G.F. Reiter,
Phys. Rev. Lett. 83, 3956 (1999); Y. Paltiel, et al., Nature {\bf 403} 398 (2000); V. V. Chabanenko, et al., Journ. of
Appl. Phys., {\bf 88}, 5875 (2000).
\bibitem{Zebouni} N.H. Zebouni, A. Venkataram,  G.N.Rao, C.G. Grenier,
J.M. Reynolds, Phys.Rev.Lett. {\bf 13}, 606 (1964).
%\bibitem{Xiao} Z.L. Xiao, E.Y. Andrei, P. Shuk, M. Greenblatt, Phys. Rev. Lett. {\bf 86}, 2431 (2001).
\bibitem{Leblank} M.A.R. Leblanc, F.L. Vernon, Physics Letters {\bf 13}, 291 (1964).
\bibitem{Geim} A.K. Geim, et al., Nature {\bf 407}, 55 (2000).
\bibitem{Suhl} H. Suhl, Phys.Rev.Lett. {\bf 14}, 226 (1965).
\bibitem{Nowak} E.R Nowak, et al., Phys.Rev.B. 55, 11702 (1997).
\bibitem{Coffey} H.T. Coffey, Cryogenics {\bf 7} 73 (1967).
\bibitem{Kopnin} N.V. Kopnin. Pis'ma v ZhETF {\bf 27}, 417 (1978); G. Baym, E. Chandler. J. Low Temp. Phys. {\bf 50}, 57 (1983); E.B. Sonin, V.B. Geshkenbein, A. van Otterlo, G. Blatter. Phys.
Rev. B {\bf 57}, 575 (1998); M.J. Stephen, J. Bardin. Phys. Rev. Lett. {\bf 14} 112 (1965); G.E. Volovik, Pis'ma v ZhETF {\bf 65} 201 (1997); E. M.W. Coffey, Phys. Rev. B {\bf 49} 9774 (1994).
\bibitem{Rose} J.I. Gittleman, B. Rosenblum. Journ. of Appl. Phys. {\bf 39} 2617 (1968).

\end{thebibliography}
\end{document}